\documentclass[aps,showpacs,prb,reprint,superscriptaddress,longbibliography]{revtex4-1}

\usepackage[colorlinks=true,linkcolor=blue,citecolor=blue,urlcolor=blue]{hyperref}

\usepackage{hyperref}
\usepackage{setspace}
\usepackage{graphicx}
\usepackage{amsmath}
\usepackage{color}
\usepackage{amssymb}
\usepackage{verbatim}
\usepackage{latexsym}
\usepackage{enumerate}
\usepackage{bm}
\usepackage{pbox}
\usepackage{multirow}

\begin{document}

\title{Lattice dynamics and electron-phonon coupling calculations using
non-diagonal supercells}

\author{Jonathan H. \surname{Lloyd-Williams}}
\email{jhl50@cam.ac.uk}
\affiliation{TCM Group, Cavendish Laboratory, University of Cambridge,
  J.\ J.\ Thomson Avenue, Cambridge CB3 0HE, United Kingdom}

\author{Bartomeu \surname{Monserrat}}
\email{bm418@cam.ac.uk}
\affiliation{TCM Group, Cavendish Laboratory, University of Cambridge,
  J.\ J.\ Thomson Avenue, Cambridge CB3 0HE, United Kingdom}
\affiliation{Department of Physics and Astronomy, Rutgers University,
  Piscataway, New Jersey 08854-8019, USA}

\date{\today}

\begin{abstract}
We study the direct calculation of total energy derivatives for lattice
dynamics and electron-phonon coupling calculations using supercell matrices 
with non-zero off-diagonal elements.
We show that it is possible to determine the response of a periodic system to a 
perturbation characterized by a wave vector with reduced fractional coordinates 
$(m_1/n_1,m_2/n_2,m_3/n_3)$ using a supercell containing a number of primitive 
cells equal to the least common multiple of $n_1$, $n_2$, and $n_3$.
If only diagonal supercell matrices are used, a supercell containing $n_1n_2n_3$
primitive cells is required.
We demonstrate that the use of non-diagonal supercells significantly reduces the
computational cost of obtaining converged zero-point energies and phonon
dispersions for diamond and graphite.
We also perform electron-phonon coupling calculations using the direct method
to sample the vibrational Brillouin zone with grids of unprecedented size,
which enables us to investigate the convergence of the zero-point
renormalization to the thermal and optical band gaps of diamond. 
\end{abstract}


\maketitle

\section{Introduction}

The experimental study of condensed matter usually involves measuring the 
response of a system to some external perturbation.
Many properties of materials can be studied theoretically by the calculation of 
derivatives of the total energy with respect to applied perturbations, such as 
force constants, elastic constants, Born effective charges, and piezoelectric 
constants~\cite{Martin_book}.
First principles methods have been successfully used to study the response of 
a wide range of systems to a variety of perturbations~\cite{Kunc_1982,Niu_1998,
Marzari_2000,Baroni_2001,Joyce_2007,Ambrosetti_2014}, complementing or 
explaining experimental discoveries, and predicting novel properties and 
behavior.

The response of periodic systems to perturbations characterized by a wave
vector can be calculated using the direct method~\cite{Yin_1980,Fleszar_1985} 
or perturbative methods~\cite{Baroni_1987,Giannozzi_1991,Gonze_1997}.
The direct method relies on freezing a perturbation into the system and 
calculating the total energy derivatives using a finite difference approach.
The result is a transparent formalism, but only perturbations commensurate with 
the simulation cell can be calculated exactly.
This presents some difficulties, for example, quantities derived from 
electron-phonon coupling matrix elements require a fine sampling of the
vibrational Brillouin zone (BZ)~\cite{Giustino_2007,Giustino_2010} and
converged results are typically not obtainable using simulation cells 
of tractable sizes. 
Perturbative methods can access perturbations at an arbitrary wave vector using 
a single primitive cell, and therefore have been the method of choice for the 
vast majority of calculations, from phonon dispersions~\cite{Giannozzi_1991} 
and electron-phonon coupling~\cite{Kong_2001} to spin 
fluctuations~\cite{Savrasov_1998}.

The simplicity of the direct method means that it typically plays a central 
role in early calculations in a given area. 
For example, it was used in the first phonon calculations for materials beyond 
$sp$-bonded metals~\cite{Chadi_1976}, and the only available electron-phonon 
coupling calculations using many-body perturbation theory rely on this 
approach~\cite{Lazzeri_2008,Gruneis_2009,Faber_2011,Yin_2013,Antonius_2014}. 
The direct method is also readily extendable to situations where large 
distortions are required, as the energy is found at all orders. 
It is therefore desirable to reduce the computational cost and consequently
extend the range of applicability of the direct method. 

In this paper, we prove that in order to calculate the response of a periodic 
system to a perturbation at a wave vector with reduced fractional coordinates 
$(m_1/n_1,m_2/n_2,m_3/n_3)$, it is only necessary to consider a supercell 
containing a number of primitive cells equal to the least common multiple (LCM) 
of $n_1$, $n_2$, and $n_3$.
This is accomplished by utilizing supercell matrices containing non-zero 
off-diagonal elements. 
For example, the sampling of the vibrational BZ with a uniform grid of size 
$N \times N \times N$ can be accomplished with supercells containing at most 
$N$ primitive cells. 
In contrast, the size of the largest supercell that may need to be considered 
scales cubically with the linear size of the BZ grid when only using diagonal 
supercell matrices.

We find that the use of non-diagonal supercell matrices reduces the 
computational cost of obtaining converged zero-point energies and phonon
dispersions for diamond and graphite by over an order of magnitude.
It also enables us to perform electron-phonon coupling calculations using the
direct method with BZ grids of unprecedented size.
In particular, we investigate the convergence with respect to the number of
points used to sample the vibrational BZ of the zero-point renormalization to
the thermal and optical band gaps of diamond, a problem that has previously 
been considered challenging for the direct approach due to the prohibitive 
computational cost of using simulation cells containing sufficient numbers of
primitive cells.

The paper is organized as follows:
We introduce the use of non-diagonal supercell matrices to access perturbations
at a given wave vector in Sec.~\ref{sec:off_diag}. 
We describe the computational details of our calculations in
Sec.~\ref{sec:comput}.
We illustrate the utility of our approach in the context of first principles
lattice dynamics in Sec.~\ref{sec:latt_dyn} and in relation to electron-phonon
coupling calculations in Sec.~\ref{sec:el_ph}. 
Our conclusions are drawn in Sec.~\ref{sec:conclusions}.

\section{Supercells and k-point sampling}\label{sec:off_diag}

\subsection{Supercell matrices}\label{sub:smat}

A simulation cell that contains multiple primitive cells of a given crystal 
lattice is known as a supercell and is itself the unit cell of a superlattice, 
whose basis vectors are constructed by taking linear combinations of the 
primitive lattice basis vectors with integer coefficients~\cite{Santoro_1972}.
This can be expressed algebraically as
\begin{equation}\label{eq:superlattice}
\begin{pmatrix}
\mathbf{a}_{\text{s}_1}\\
\mathbf{a}_{\text{s}_2}\\
\mathbf{a}_{\text{s}_3}
\end{pmatrix}
=
\begin{pmatrix}
S_{11}&S_{12}&S_{13}\\
S_{21}&S_{22}&S_{23}\\
S_{31}&S_{32}&S_{33}
\end{pmatrix}
\begin{pmatrix}
\mathbf{a}_{\text{p}_1}\\
\mathbf{a}_{\text{p}_2}\\
\mathbf{a}_{\text{p}_3}
\end{pmatrix}
\,,
\end{equation}
where $\mathbf{a}_{\text{s}_i}$ are the superlattice basis vectors, 
$\mathbf{a}_{\text{p}_i}$ are the primitive lattice basis vectors, and 
$S_{ij}\in\mathbb{Z}$.
The supercell contains $|S|$ primitive cells and we refer to the matrix $S$ as 
the supercell matrix.
For the purposes of brevity, we shall henceforth refer to supercells generated 
by diagonal supercell matrices as \textit{diagonal supercells} and those 
generated by non-diagonal supercell matrices as 
\textit{non-diagonal supercells}.

The set of wave vectors that describe plane waves with the same periodicity as 
the primitive lattice define the reciprocal primitive lattice with basis 
vectors
\begin{equation}
\begin{pmatrix}
\mathbf{b}_{\text{p}_1}\\
\mathbf{b}_{\text{p}_2}\\
\mathbf{b}_{\text{p}_3}
\end{pmatrix}
=2\pi
\begin{pmatrix}
\mathbf{a}_{\text{p}_1}\\
\mathbf{a}_{\text{p}_2}\\
\mathbf{a}_{\text{p}_3}
\end{pmatrix}^{-\text{T}}\,,
\end{equation}
and the set of wave vectors that describe plane waves with the same periodicity 
as the superlattice define the reciprocal superlattice with basis vectors
\begin{equation}
\begin{pmatrix}
\mathbf{b}_{\text{s}_1}\\
\mathbf{b}_{\text{s}_2}\\
\mathbf{b}_{\text{s}_3}
\end{pmatrix}
=
\begin{pmatrix}
\bar{S}_{11}&\bar{S}_{12}&\bar{S}_{13}\\
\bar{S}_{21}&\bar{S}_{22}&\bar{S}_{23}\\
\bar{S}_{31}&\bar{S}_{32}&\bar{S}_{33}
\end{pmatrix}
\begin{pmatrix}
\mathbf{b}_{\text{p}_1}\\
\mathbf{b}_{\text{p}_2}\\
\mathbf{b}_{\text{p}_3}
\end{pmatrix}\,,
\end{equation}
where $\bar{S}_{ij}=(S^{-1})_{ji}$.
An arbitrary $\mathbf{k}$-point can be expressed in terms of both the
reciprocal primitive lattice basis vectors and reciprocal superlattice basis 
vectors, and these fractional coordinates are related by
\begin{equation}
\begin{pmatrix}
k_{\text{s}_1}\\
k_{\text{s}_2}\\
k_{\text{s}_3}
\end{pmatrix}
=
\begin{pmatrix}
S_{11}&S_{12}&S_{13}\\
S_{21}&S_{22}&S_{23}\\
S_{31}&S_{32}&S_{33}
\end{pmatrix}
\begin{pmatrix}
k_{\text{p}_1}\\
k_{\text{p}_2}\\
k_{\text{p}_3}
\end{pmatrix}\,.
\end{equation}
If the reciprocal superlattice fractional coordinates are all integers, 
perturbations characterized by the wave vector $\mathbf{k}$ 
are commensurate with the supercell generated by $S$.

There are a finite number of unique superlattices whose supercells contain
a given number of primitive cells, but there are an infinite number of sets of 
basis vectors that can be used to describe each superlattice.
Two different supercell matrices $S$ and $S'$ generate different bases for
the same superlattice if $S'$ can be reduced to $S$ by elementary unimodular 
row operations~\cite{Hart_2008}, which consist of the following:
\begin{itemize}
\item Adding an integer multiple of one row of the matrix to another row.
\item Interchanging two rows of the matrix.
\item Multiplying a row of the matrix by $-1$.
\end{itemize}
The canonical form for such operations is the upper-triangular Hermite normal
form (HNF):
\begin{equation}\label{eq:HNF}
\begin{pmatrix}
S_{11}&S_{12}&S_{13}\\0&S_{22}&S_{23}\\0&0&S_{33}
\end{pmatrix},\,
\end{equation}
with $0 \leq S_{12} < S_{22}$ and $0 \leq S_{13}, S_{23} < S_{33}$.
This means that all inequivalent supercell matrices can be written in the form
given by Eq.~(\ref{eq:HNF}).
Note that the product $S_{11}S_{22}S_{33}$ fixes the determinant $|S|$ and
therefore the number of primitive cells contained within the supercell.

\subsection{Commensurate supercells}\label{sub:proof}

We now show that a $\mathbf{k}$-point with fractional coordinates
\begin{equation}
\begin{pmatrix}
k_{\text{p}_{1}}\\
k_{\text{p}_{2}}\\
k_{\text{p}_{3}}
\end{pmatrix}
=
\begin{pmatrix}
\frac{m_{1}}{n_{1}}\\
\frac{m_{2}}{n_{2}}\\
\frac{m_{3}}{n_{3}}
\end{pmatrix}\,,
\end{equation}
where $0 \leq k_{\text{p}_{1}}, k_{\text{p}_{2}}, k_{\text{p}_{3}} < 1$ and $m_{1}/n_{1}$,
$m_{2}/n_{2}$, and $m_{3}/n_{3}$ are reduced fractions, is commensurate with a 
supercell containing $l_{123}$ primitive cells, where $l_{123}$ is the LCM of 
$n_{1}$, $n_{2}$, and $n_{3}$.
That is to say, we are able to solve the equations
\begin{align}
k_{\text{s}_{1}}&=\frac{S_{11}m_{1}}{n_{1}}+\frac{S_{12}m_{2}}{n_{2}}+
\frac{S_{13}m_{3}}{n_{3}}\label{eq:k1}\\
k_{\text{s}_{2}}&=\frac{S_{22}m_{2}}{n_{2}}+\frac{S_{23}m_{3}}{n_{3}}\label{eq:k2}\\
k_{\text{s}_{3}}&=\frac{S_{33}m_{3}}{n_{3}}\label{eq:k3}
\end{align}
for integer $k_{\text{s}_{1}}$, $k_{\text{s}_{2}}$, and $k_{\text{s}_{3}}$ with 
$S_{11}S_{22}S_{33}=l_{123}$ and $S_{12}$, $S_{13}$, and $S_{23}$ satisfying the
conditions stated above. 
The proof that follows uses properties of complete and reduced residue 
systems, which are detailed in the Appendix.

We trivially solve Eq.~(\ref{eq:k3}) by setting $S_{33}=n_{3}$.
We solve Eq.~(\ref{eq:k2}) by setting $S_{22}=n_{2}/g_{23}$ and $S_{23}=pn_{3}/
g_{23}$, where $g_{23}$ is the greatest common divisor (GCD) of $n_{2}$ and 
$n_{3}$, and $p$ is a non-negative integer, which results in
\begin{equation}\label{eq:proof1}
k_{\text{s}_{2}}=\frac{m_{2}}{g_{23}}+\frac{pm_{3}}{g_{23}}.
\end{equation}
The condition $S_{23} < S_{33}$ requires that $p < g_{23}$.
If $g_{23}=1$, $p=0$, and if $g_{23}>1$, $m_{3}\,\text{mod}\,g_{23}$ is a 
generator for the additive group of integers modulo $g_{23}$.
This follows since $m_{3}/n_{3}$ is a reduced fraction and $g_{23}$ divides 
$n_{3}$.
We can therefore always choose $p<g_{23}$ such that $(m_{2}+pm_{3})\,\text{mod}\,
g_{23}=0$ and obtain an integer solution for $k_{\text{s}_{2}}$.

We now consider Eq.~(\ref{eq:k1}) and proceed by setting $S_{11}=g_{123}n_{1}/
(g_{12}g_{31})$, $S_{12}=qg_{123}n_{2}/(g_{12}g_{23})$, and $S_{13}=rg_{123}n_{3}/
(g_{31}g_{23})$, where $g_{12}$ is the GCD of $n_{1}$ and $n_{2}$, $g_{31}$ is
the GCD of $n_{3}$ and $n_{1}$, $g_{123}$ is the GCD of $n_{1}$, $n_{2}$, and
$n_{3}$, and $q$ and $r$ are non-negative integers, which results in
\begin{equation}\label{eq:proof2}
\begin{split}
k_{\text{s}_{1}}=&\phantom{+}\frac{g_{123}}{g_{31}g_{23}}\left[
\frac{(g_{23}/g_{123})m_{1}}{g_{12}/g_{123}}+
\frac{q(g_{31}/g_{123})m_{2}}{g_{12}/g_{123}}
\right]\\
&+\frac{rg_{123}m_{3}}{g_{31}g_{23}}\,.
\end{split}
\end{equation}
The condition $S_{12}<S_{22}$ requires that $q<g_{12}/g_{123}$ and the 
condition $S_{13}<S_{33}$ requires that $r<g_{31}g_{23}/g_{123}$.
If $g_{12}/g_{123}=1$, $q=0$, and if $g_{12}/g_{123}>1$, $m_{2}\,\text{mod}\,
(g_{12}/g_{123})$ is an element of the multiplicative group of integers modulo
$g_{12}/g_{123}$.
This follows since $m_{2}/n_{2}$ is a reduced fraction, $g_{12}$ divides 
$n_{2}$, and $g_{123}$ divides $g_{12}$.
$(g_{31}/g_{123})\,\text{mod}\,(g_{12}/g_{123})$ is also an element of the 
multiplicative group of integers modulo $g_{12}/g_{123}$.
This follows since $g_{123}$ is the GCD of $g_{12}$ and $g_{31}$, which means 
that $(g_{31}/g_{123})/(g_{12}/g_{123})$ is a reduced fraction.
The product $(g_{31}/g_{123})m_{2}\,\text{mod}\,(g_{12}/g_{123})$ is a generator 
for the additive group of integers modulo $g_{12}/g_{123}$ and we can therefore 
always choose $q<g_{12}/g_{123}$ such that $((g_{23}/g_{123})m_{1}+
q(g_{31}/g_{123})m_{2})\,\text{mod}\,(g_{12}/g_{123})=0$.
Both of these cases lead to
\begin{equation}\label{eq:proof4}
k_{\text{s}_{1}}=\frac{z}{g_{31}g_{23}/g_{123}}+
\frac{rm_{3}}{g_{31}g_{23}/g_{123}}\,,
\end{equation}
where $z$ is an integer.
If $g_{31}g_{23}/g_{123}=1$, $r=0$, and if $g_{31}g_{23}/g_{123}>1$, $m_{3}\,
\text{mod}\,(g_{31}g_{23}/g_{123})$ is a generator for the additive group of 
integers modulo $g_{31}g_{23}/g_{123}$.
This follows since $m_{3}/n_{3}$ is a reduced fraction, $g_{31}$ and $g_{23}$
both divide $n_{3}$, and $g_{123}$ divides both $g_{31}$ and $g_{23}$.
We can therefore always choose $r<g_{31}g_{23}/g_{123}$ such that $(z+rm_{3})\,
\text{mod}\,(g_{31}g_{23}/g_{123})=0$ and obtain an integer solution for 
$k_{\text{s}_{1}}$.

Given the choice of $S_{11}$, $S_{22}$, and $S_{33}$ stated above, the number 
of primitive cells contained within the supercell is $g_{123}n_{1}n_{2}n_{3}/
g_{12}g_{23}g_{31}=l_{123}$.
In particular, we can access all $\mathbf{k}$-points
on a uniform $N \times N \times N$ grid by considering supercells containing at 
most $N$ primitive cells.

\subsection{Two-dimensional example}\label{sub:2d}

\begin{figure}
\centering
\includegraphics[width=\linewidth]{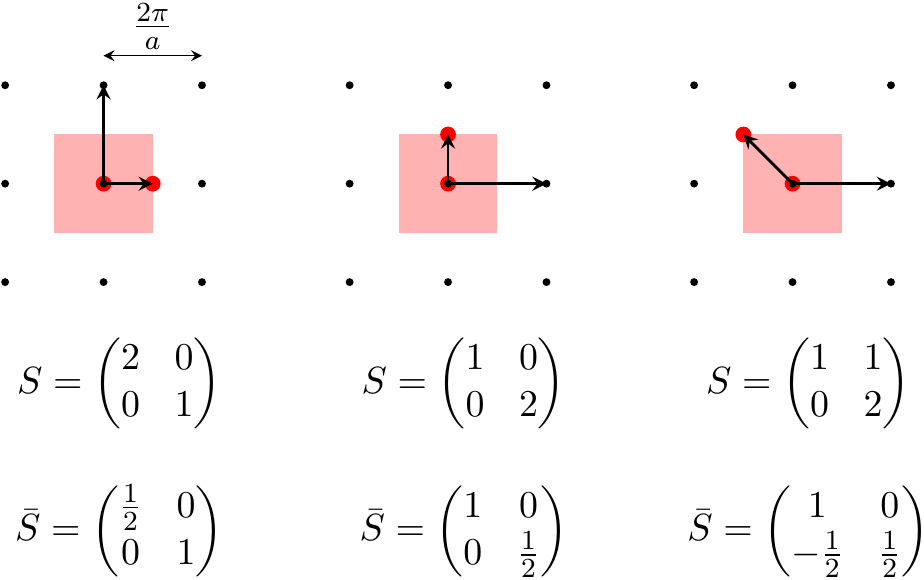}
\caption{(color online) Example of the use of non-diagonal supercells to access 
all points on a $2\times 2$ grid sampling the BZ (red shaded area) of a square 
lattice (black dots). The black arrows indicate the reciprocal superlattice 
vectors and the red dots indicate the points in the first BZ that are 
accessible to them.}
\label{fig:2d_example}
\end{figure}

We now describe a two-dimensional example of the use of non-diagonal supercells
to access any point on a $2\times 2$ grid sampling the BZ of a square lattice 
with spacing $a$. 
The reciprocal lattice is also square, but with lattice parameter $2\pi/a$.

In Fig.~\ref{fig:2d_example}, we show the reciprocal lattice together with the 
first BZ (shaded red area). 
The points of a $2\times 2$ grid on the BZ have fractional coordinates $(0,0)$, 
$(\frac{1}{2},0)$, $(0,\frac{1}{2})$, and $(\frac{1}{2},\frac{1}{2})$. 
The centre of the BZ, $(0,0)$, is commensurate with a primitive cell. 
Diagonal supercells with $|S|=2$ may be used to access the points 
$(\frac{1}{2},0)$ and $(0,\frac{1}{2})$, as shown in Fig.~\ref{fig:2d_example}.
The point $(\frac{1}{2},\frac{1}{2})$ cannot be accessed with a diagonal 
supercell of size $|S|=2$, and instead the smallest diagonal supercell that 
provides access to this point has size $|S|=4$. 
However, as shown in Fig.~\ref{fig:2d_example}, a non-diagonal supercell of 
size $|S|=2$ provides access to the point $(-\frac{1}{2},\frac{1}{2})$, which 
is equivalent to the point $(\frac{1}{2},\frac{1}{2})$.

\subsection{Other uses of non-diagonal supercells}

Non-diagonal supercells have been previously used for the calculation of
phonon dispersion curves along high symmetry lines using the planar force
constant method~\cite{Yin_1982,Kunc_1985}.
In this context, supercells are constructed by hand to capture the force
constants arising from finite displacements of entire planes of atoms.
Interatomic force constants may, in principle, be obtained from a set of such
interplanar force constants using a least-squares fit
procedure~\cite{Wei_1992}.
A similar procss may be carried out using a combination of supercells that 
maximizes the cutoff radius of the force constants~\cite{Heid_1998}.
In Sec.~\ref{sec:latt_dyn} below, we show how non-diagonal supercells can be 
used to directly construct the dynamical matrices required for lattice
dynamics calculations in the harmonic approximation, thus generalizing and
systematizing previous approaches. 

More widely, non-diagonal supercells have been used in real space methods 
such as quantum Monte Carlo for the study of solids from first
principles~\cite{Foulkes_2001,Drummond_2015}, or the Lanczos method for the
study of model systems~\cite{Dagotto_1994,Kent_2005}.
In these cases, non-diagonal supercells are used to construct appropriate 
simulation cells to facilitate the extrapolation of finite system size results
to the infinite system limit, as well as for the calculation of total energy
derivatives to evaluate susceptibilities.

\section{Computational details}\label{sec:comput}

\subsection{Non-diagonal supercell generation}

We now describe how we use non-diagonal supercells to perform calculations of 
total energy derivatives using the direct method in practice.
We express each $\mathbf{k}$-point of interest in reduced fractional 
coordinates and calculate $l_{123}$ to determine the size of supercell $|S|$ 
commensurate with it.
We choose the appropriate supercell matrix in HNF and then perform elementary 
unimodular row operations on it until the superlattice basis vectors are the 
shortest possible.
We have found this to reduce the total number of points required to sample the 
electronic BZ for a fixed Monkhorst-Pack~\cite{Monkhorst_1976} grid spacing 
criterion, which helps to minimize the computational cost of our first
principles calculations.
A {\sc fortran 90} program implementing this procedure is included in the
Supplemental Material~\cite{Supplemental}.

\subsection{First principles calculations}

We have studied diamond and graphite using plane wave pseudopotential density 
functional theory~\cite{Hohenberg_1964,Kohn_1965}, as implemented in version 8 
of the {\sc castep} code~\cite{CASTEP}.
We used the local density approximation~\cite{Ceperley_1980,Perdew_1981} 
to the exchange-correlation functional and an ``on-the-fly'' ultrasoft 
pseudopotential~\cite{Vanderbilt_1990} generated by {\sc castep} with valence 
states $2\text{s}^{2}2\text{p}^2$. 
We used a plane wave energy cutoff of $800$~eV and sampled the electronic BZ 
with a Monkhorst-Pack~\cite{Monkhorst_1976} grid of density $2 \pi \times 
0.03$~{\AA}$^{-1}$, which was sufficient to converge the energy differences 
between different frozen phonon configurations to better than $10^{-4}$~eV per 
atom.
We relaxed the structures at zero pressure until the forces on each atom were 
smaller than $10^{-4}$~eV/{\AA} and the components of the stress tensor were 
smaller than $10^{-4}$~GPa, which resulted in a lattice constant of 
$3.532$~{\AA} for diamond, and an in-plane lattice parameter of $2.445$~{\AA}
with $c/a=2.707$ for graphite.
These values are slightly smaller than the experimental
ones~\cite{Grenville-Wells_1958,Hanfland_1989,Zhao_1989}, which is because the 
local density approximation favors uniform charge densities and therefore tends 
to overbind.
The lattice constant of diamond is also different to that obtained with version 
7 of {\sc castep} used for Ref.~\onlinecite{Monserrat_Diamond_2014}, which is 
due to a change in the default ``on-the-fly'' pseudopotential.

\section{Lattice dynamics}\label{sec:latt_dyn}

\subsection{Formalism}

Assuming Born-von Karman periodic boundary conditions~\cite{Born_Huang} applied
to an $N_{1} \times N_{2} \times N_{3}$ array of primitive cells, the central
question of first principles lattice dynamics in the harmonic
approximation~\cite{Wallace} is how to determine the so-called dynamical
matrix at each $\mathbf{k}$-point on an $N_{1} \times N_{2} \times N_{3}$
grid sampling the vibrational BZ.
The dynamical matrix is defined as
\begin{equation}\label{eq:dyn_mat}
D_{ij}(\alpha\beta\,|\,\mathbf{k})=
\frac{1}{\sqrt{m_{\alpha}m_{\beta}}}
\sum_{\mathbf{R}_{p}}
\Phi_{ij}(\alpha\beta\,|\,\mathbf{R}_{p})
\text{e}^{-\text{i}\mathbf{k}\cdot\mathbf{R}_{p}}\,,
\end{equation}
where Latin indices label Cartesian coordinates, Greek indices label the
atoms within a primitive cell, $m_{\alpha}$ is the mass of atom $\alpha$,
$\mathbf{R}_{p}$ are the position vectors of the primitive cells that make
up the simulation cell, and
\begin{equation}
\Phi_{ij}(\alpha\beta\,|\,\mathbf{R}_{p}-\mathbf{R}_{p'})=
\frac{\partial^{2}E_{\text{BO}}}
{\partial u_{i}(\alpha\,|\,\mathbf{R}_{p})
\partial u_{j}(\beta\,|\,\mathbf{R}_{p'})}\,,
\end{equation}
where $E_{\text{BO}}$ is the Born-Oppenheimer (BO) potential energy
surface~\cite{Born_1927} and $u_{i}(\alpha\,|\,\mathbf{R}_{p})$ is the $i$th
component of the displacement from its equilibrium position of the $\alpha$th
atom in the primitive cell located at $\mathbf{R}_{p}$, is the matrix of
interatomic force constants.
The eigenvectors of the dynamical matrix can be used to rewrite the harmonic
vibrational Hamiltonian in terms of normal coordinates $q_{n\mathbf{k}}$,
where $n$ is the phonon branch index.
The Hamiltonian then takes the form of a sum of terms corresponding to
non-interacting simple harmonic oscillators with frequencies $\omega_{n
\mathbf{k}}$ equal to the square root of the eigenvalues of the dynamical
matrix.
The resulting vibrational eigenstates can be found analytically.

The matrix of force constants decays with distance between primitive cells and
consequently it is possible to obtain an excellent approximation to the exact
dynamical matrix at an arbitrary wave vector if the simulation cell is
sufficiently large~\cite{Yin_1982}.
Therefore, the standard approach is to determine the dynamical matrix at each
symmetry-inequivalent $\mathbf{k}$-point on the $N_{1} \times N_{2} \times
N_{3}$ grid (typically referred to as the \textit{coarse} grid), construct the 
matrix of force constants corresponding to the $N_{1} \times N_{2} \times N_{3}$
array of primitive cells using the inverse of Eq.~(\ref{eq:dyn_mat}), and then 
calculate phonon frequencies and atomic displacement patterns at a large number 
of $\mathbf{k}$-points, which can be used to compute structural, vibrational,
and thermodynamic properties of the system.
The first step of this process can be achieved either by using density
functional perturbation theory~\cite{Baroni_2001} to determine the linear
response of the charge density to an atomic displacement characterized by a
wave vector $\mathbf{k}$, or by directly calculating the matrix of force
constants using a supercell commensurate with $\mathbf{k}$ and performing the
Fourier transform given by Eq.~(\ref{eq:dyn_mat}).
The direct approach takes advantage of the fact that the dynamical matrix is
exact at a given $\mathbf{k}$-point, in the sense that it is equal to its
infinite system counterpart, if it is constructed using force constants
calculated with a supercell commensurate with the wave vector $\mathbf{k}$.
It is convenient to restrict the exact calculation of the dynamical matrix to
$\mathbf{k}$-points in the irreducible wedge of the BZ (IBZ) and obtain it at
all other points by exploiting the crystal symmetries~\cite{Maradudin_1968}.
Numerical noise can slightly break the symmetry of the dynamical matrix
elements at each wave vector but this is corrected by symmetrizing them with
respect to the point group operations of the crystal that leave the wave vector 
unchanged~\cite{Worlton_1972}.

\subsection{Results}

\begin{figure}
\centering
\includegraphics[width=\linewidth]{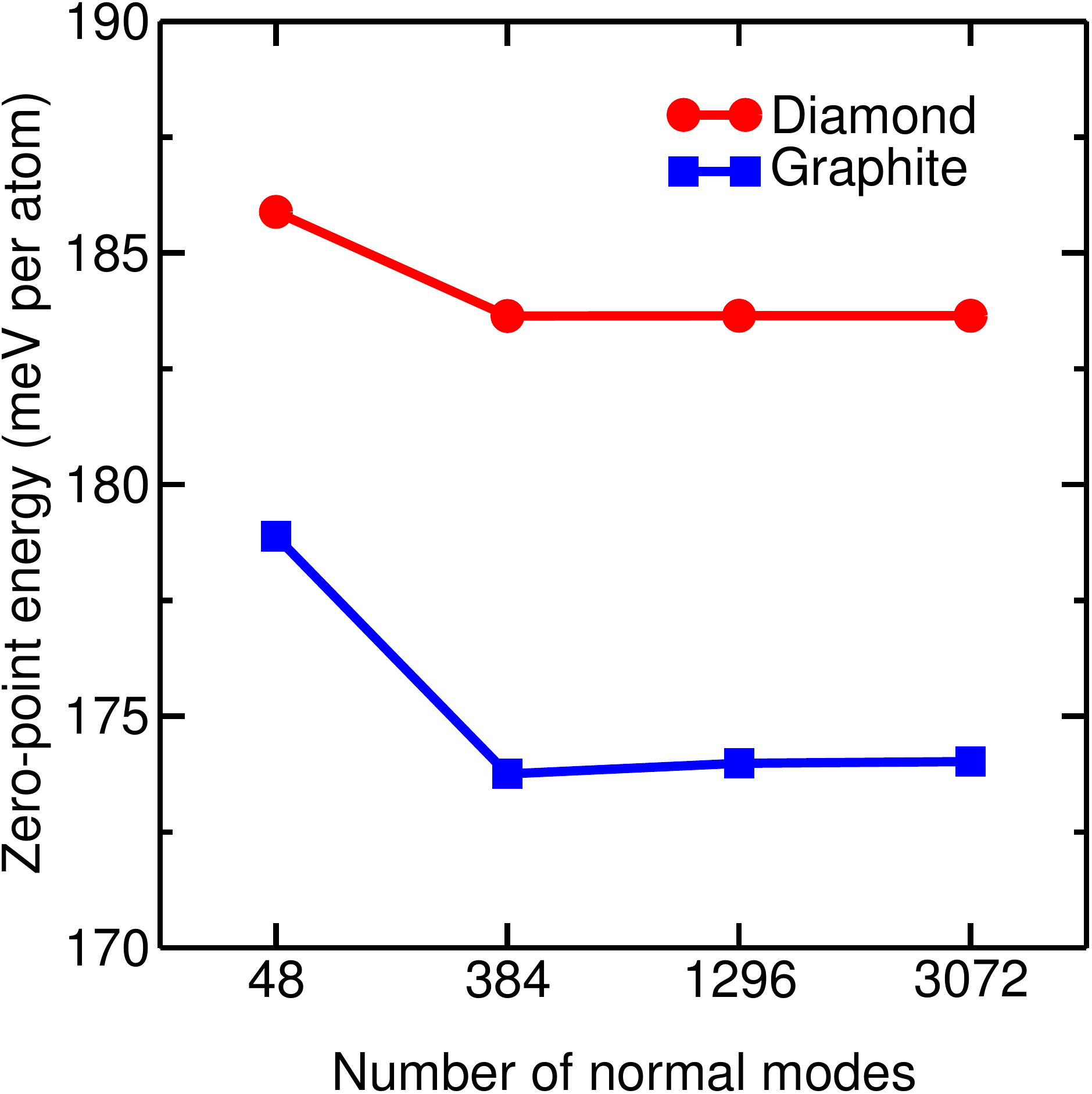}
\caption{(color online) Zero-point energy as a function of the number of normal 
modes (coarse grid size) for diamond (red circles) and graphite (blue 
squares). The solid lines are a guide to the eye.}
\label{fig:harmonic}
\end{figure}

In Fig.~\ref{fig:harmonic}, we show the convergence with respect to the number 
of normal modes included on the coarse grid used to sample the vibrational BZ 
of the zero-point energy for diamond and graphite.
The zero-point energy is converged to better than $0.1$~meV per atom using a $4 
\times 4 \times 4$ grid with $384$ normal modes for diamond and a $6 \times 6 
\times 3$ grid with $1,296$ normal modes for graphite.

\begin{figure}
\centering
\includegraphics[width=\linewidth]{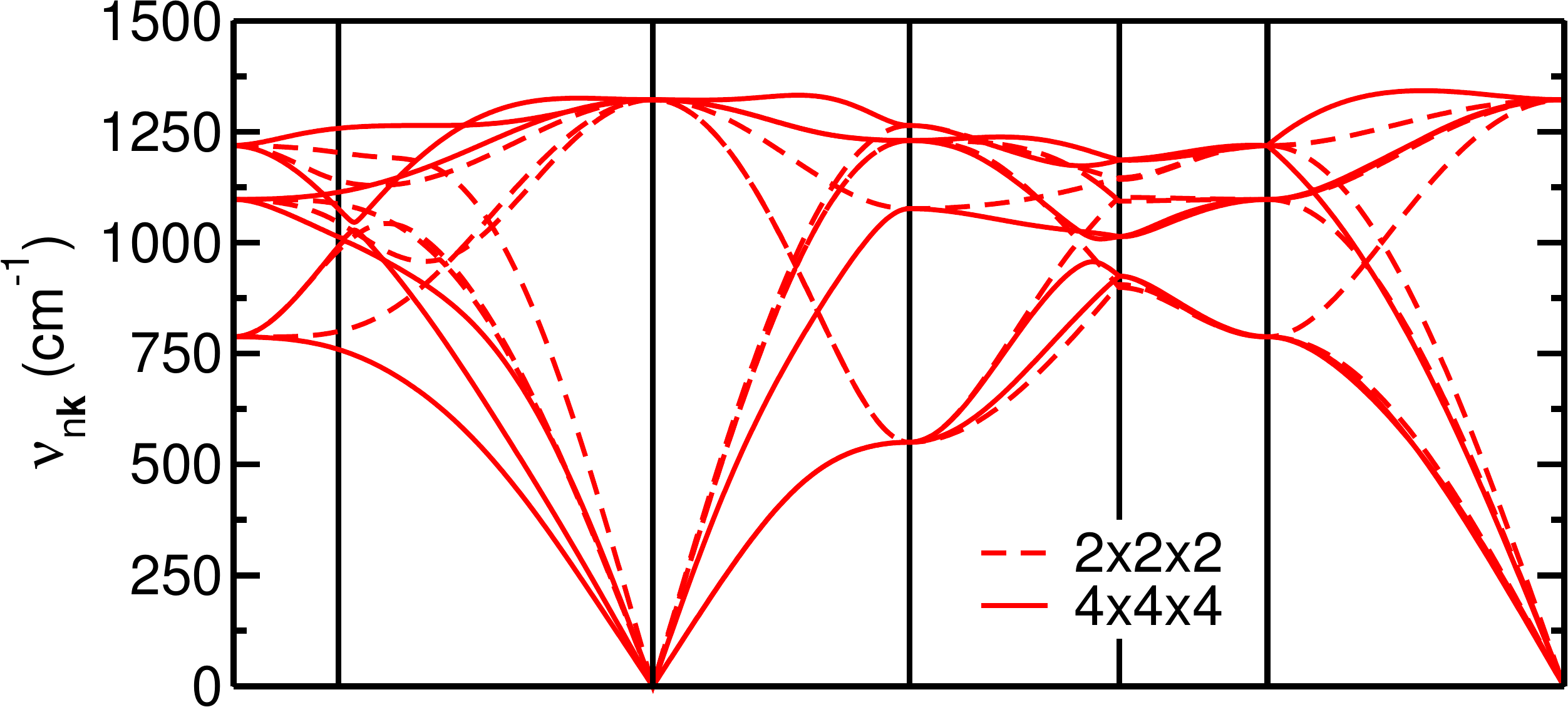}
\includegraphics[width=\linewidth]{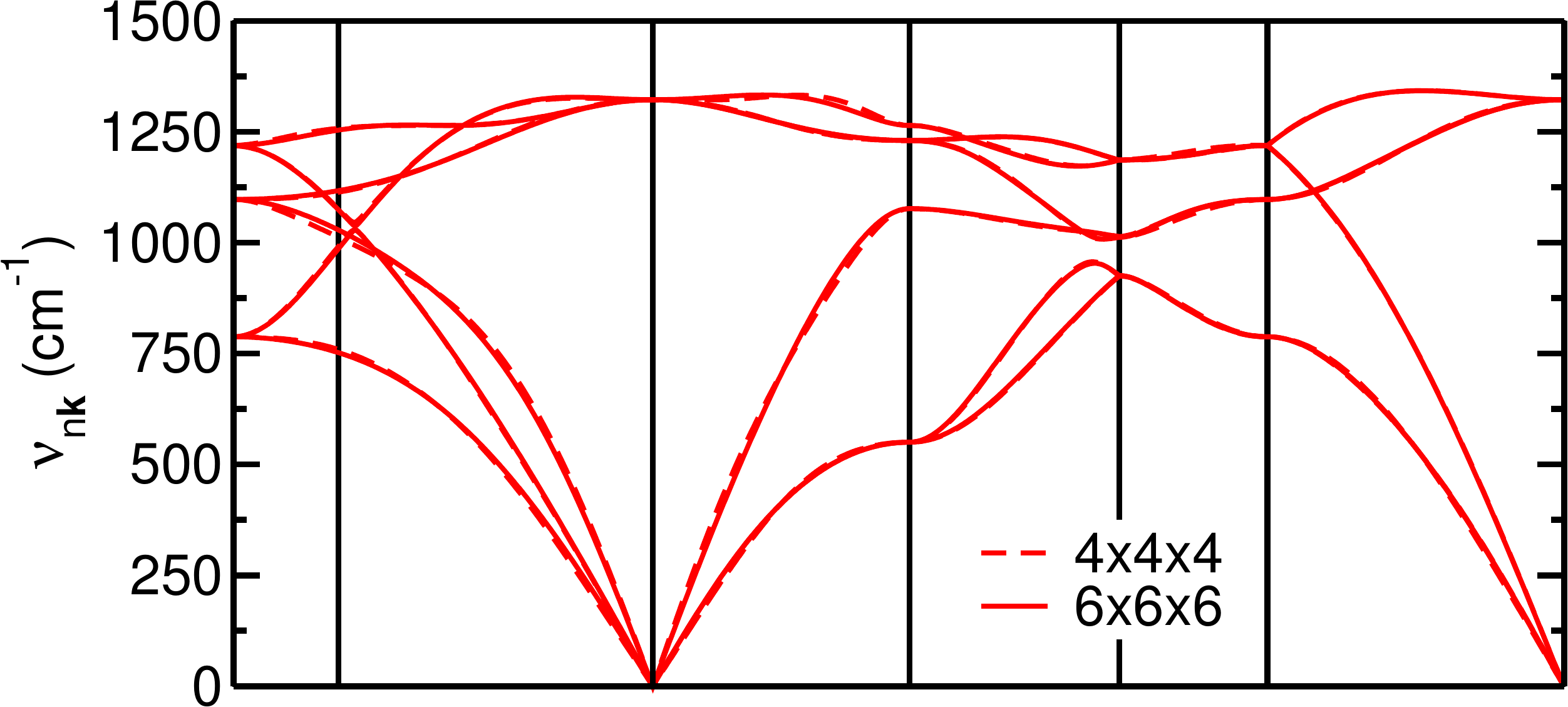}
\includegraphics[width=\linewidth]{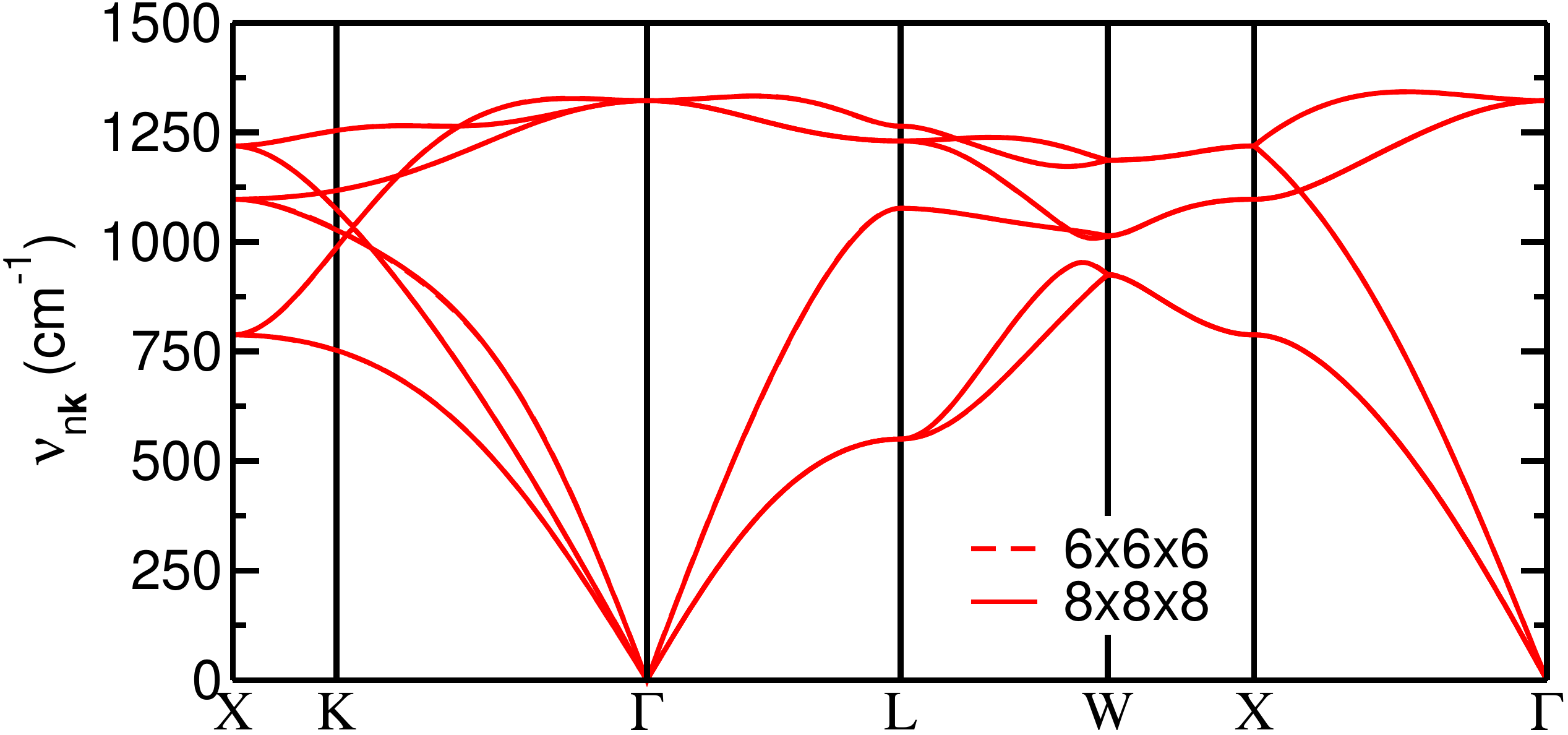}
\caption{(color online) Convergence with respect to coarse grid size of phonon 
dispersions along symmetry lines for diamond.}
\label{fig:disp_diamond}
\end{figure}

\begin{figure}
\centering
\includegraphics[width=\linewidth]{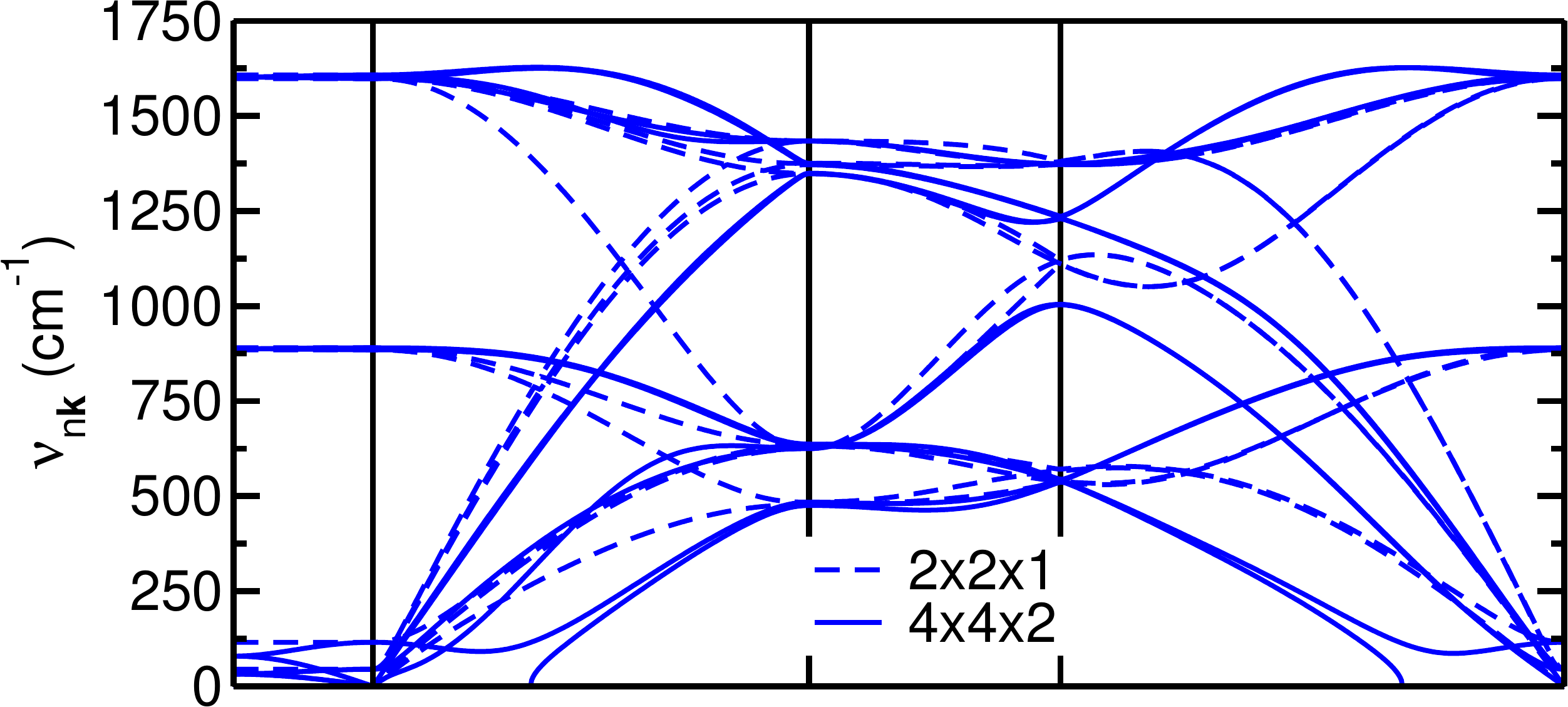}
\includegraphics[width=\linewidth]{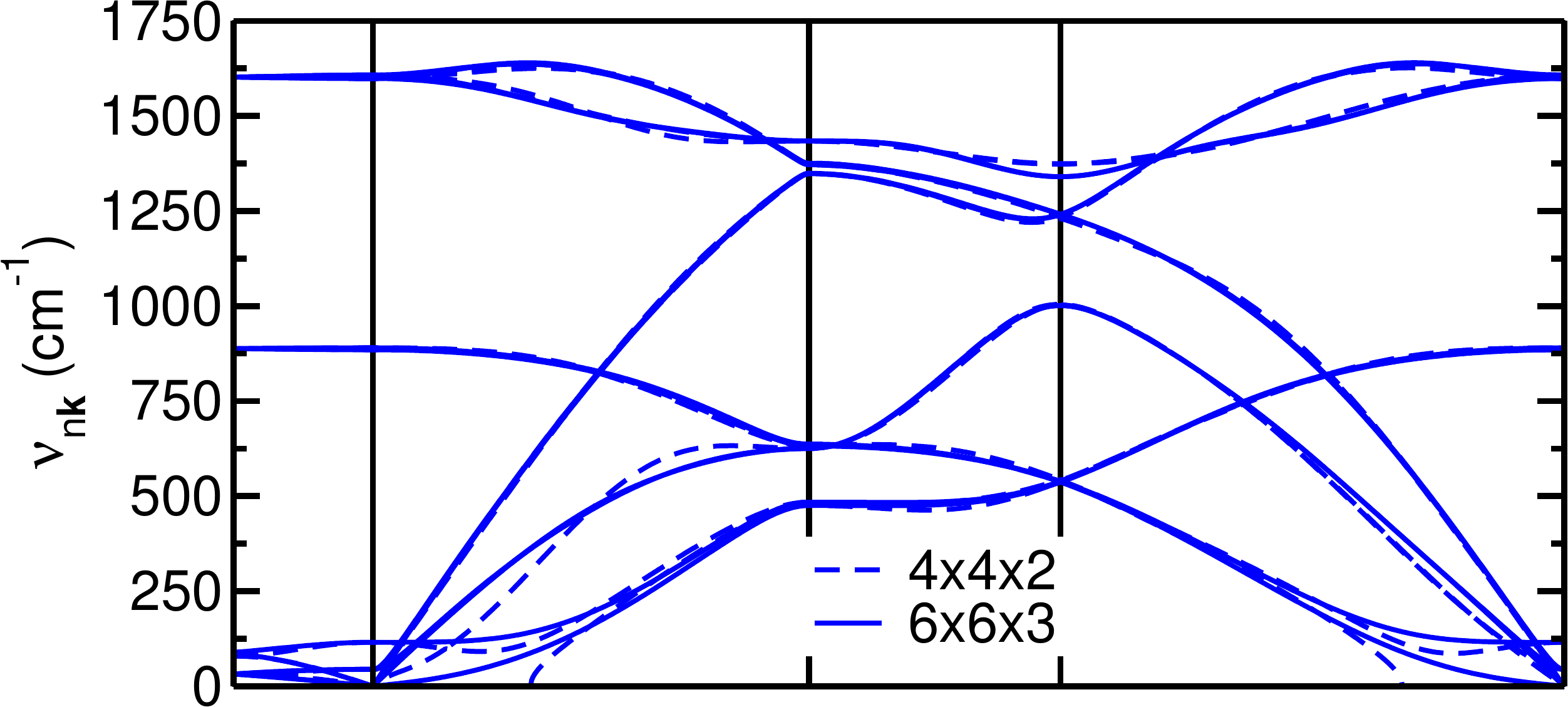}
\includegraphics[width=\linewidth]{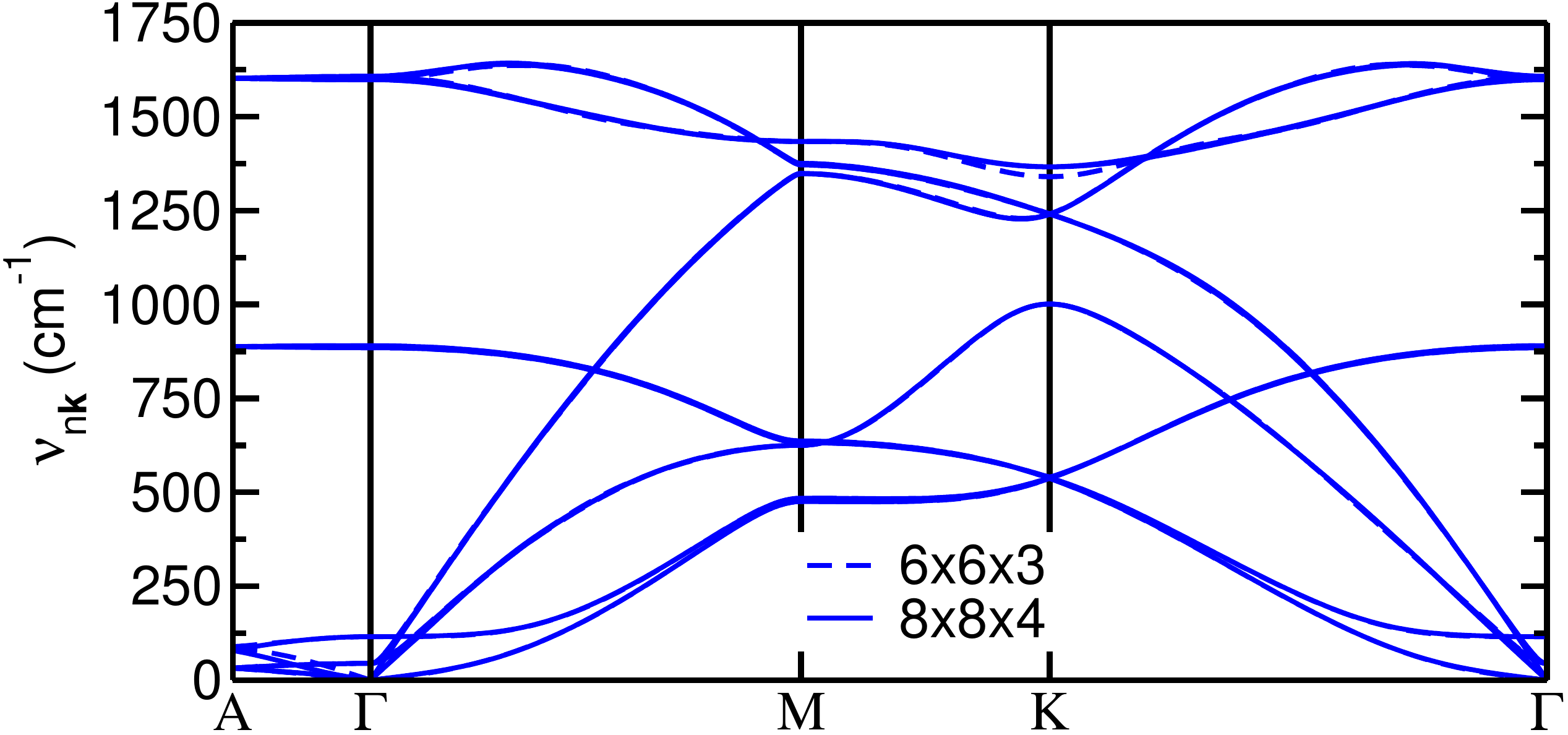}
\caption{(color online) Convergence with respect to coarse grid size of phonon 
dispersions along symmetry lines for graphite.}
\label{fig:disp_graphite}
\end{figure}

We have also investigated the convergence of phonon dispersion relations along
lines between high symmetry points in the vibrational BZ, as shown in
Figs.~\ref{fig:disp_diamond} and~\ref{fig:disp_graphite}.
With the exception of acoustic branches in the immediate vicinity of $\Gamma$,
which have negligible frequencies and are absolutely converged to better than
$5~\text{cm}^{-1}$, we find that the phonon dispersions are converged to $1-2\%$
using a $4 \times 4 \times 4$ grid for diamond and a $6 \times 6 \times 3$ grid 
for graphite.

\begin{table}
\caption{Comparison of total computational cost of calculating force constants
for diamond and graphite in order to construct the dynamical matrix at
different grid sizes when using diagonal and non-diagonal supercells.}
\label{tab:timing}
\begin{tabular}{ccc}
\hline
\hline
\multirow{2}{*}{Diamond} & \multirow{2}{*}{Grid} & Ratio of CPU time \\
& & (Diagonal : Non-diagonal) \\
\hline
& $2 \times 2 \times 2$ & $2.23$ \\
& $4 \times 4 \times 4$ & $15.3$ \\
& $6 \times 6 \times 6$ & $16.8$ \\
\hline
\multirow{2}{*}{Graphite} & \multirow{2}{*}{Grid} & Ratio of CPU time \\
& & (Diagonal : Non-diagonal) \\
\hline
& $2 \times 2 \times 1$ & $1.15$ \\
& $4 \times 4 \times 2$ & $5.29$ \\
& $6 \times 6 \times 3$ & $13.8$ \\
\hline
\hline
\
\end{tabular}
\end{table}

In Table~\ref{tab:timing}, we compare the computational cost when using 
diagonal and non-diagonal supercells of obtaining the dynamical matrix at all 
$\mathbf{k}$-points in the IBZ for diamond and graphite, respectively.
The use of a non-diagonal supercell can reduce the symmetry of the
superlattice, which determines the number of atomic displacements required to 
obtain the full matrix of force constants.
However, the ability to perform the necessary calculations at smaller systems
sizes when using non-diagonal supercells results in a lower overall
computational cost.
More detailed timing information is included in the Supplemental
Material~\cite{Supplemental}.

Considering diamond with a $4 \times 4 \times 4$ coarse grid, the most
expensive calculation is that required to determine the dynamical matrix at the
fractional $\mathbf{k}$-point $(1/4,1/2,-1/4)$.
This can be achived by employing a diagonal $4 \times 2 \times 4$ supercell
containing $32$ primitive cells.
Therefore, the full $4 \times 4 \times 4$ supercell does not actually need to
be constructed, as it is computationally cheaper to calculate the dynamical
matrices at all points in the IBZ using multiple diagonal supercells.
When using non-diagonal supercells, the largest supercells required
contain just $4$ primitive cells, and the overall speedup is greater than a
factor of ten.

Considering graphite with a $6 \times 6 \times 3$ coarse grid, the most
expensive calculation is that required to determine the dynamical matrix at the 
fractional $\mathbf{k}$-point $(1/6,1/6,1/3)$.
In this case, a diagonal $6 \times 6 \times 3$ supercell containing $108$
primitive cells must be constructed, which generates the force constants
required to calculate the dynamical matrices at all points in the IBZ that are
not present on the smaller grids.
When using multiple non-diagonal supercells, the largest supercells required
contain just $6$ primitive cells, and the overall speedup is greater than a
factor of ten.

For both the zero-point energy and phonon dispersion relations, converged
results are obtained using a $4 \times 4 \times 4$ coarse grid for diamond and 
a $6 \times 6 \times 3$ coarse grid for graphite.
As shown in table~\ref{tab:timing}, the cost of performing these calculations 
is reduced by over an order of magnitude when non-diagonal supercells are used 
instead of only diagonal supercells.

The harmonic approximation relies on the assumption that the displacement of
atoms from their equilibrium positions is sufficiently small for the
BO potential energy surface to be accurately approximated by a Taylor series
expansion around the equilibrium atomic configuration that is truncated at
second order.
Therefore, it breaks down when the atomic vibrational amplitudes are large.
A number of different approaches based on the direct method have recently been
proposed for studying anharmonicity in solids from first
principles~\cite{Souvatzis_2008,Hellman_2011,Antolin_2012,Monserrat_2013,
Errea_2014}.
A common feature of these methods is that they require the sampling of the
BO potential energy surface at a large number of atomic configurations, which
is a process that may be greatly expedited by the use of non-diagonal
supercells.

\section{Electron-phonon coupling}\label{sec:el_ph}

\subsection{Formalism}

The effect of electron-phonon coupling on the band gap of a semiconductor can
be calculated by determining the change in the electronic band structure due to 
the displacement of atoms from their equilibrium 
positions~\cite{Allen_1976,Allen_1981}.
We calculate the vibrationally averaged band gap $\langle E_{\text{g}} \rangle$ 
at zero temperature in the BO approximation as
\begin{equation}\label{eq:band_gap}
\langle E_{\text{g}} \rangle =
\int\text{d}\mathbf{q}\,|\Phi(\mathbf{q})|^{2}\,E_{\text{g}}(\mathbf{q})
\,,
\end{equation}
where $\mathbf{q}$ is a collective vibrational coordinate with elements 
$q_{n\mathbf{k}}$ and $\Phi(\mathbf{q})$ is the vibrational wave function.
Within the harmonic approximation, $\Phi(\mathbf{q})$ is a product over normal
modes of simple harmonic oscillator eigenstates.
The expression in Eq.~(\ref{eq:band_gap}) can be evaluated using Monte 
Carlo sampling~\cite{Patrick_2013,Monserrat_Helium_2014}, molecular 
dynamics~\cite{Pan_2014}, path integral 
methods~\cite{Ramirez_2006,Morales_2013}, or by using a series expansion of the 
form~\cite{Allen_1976,Allen_1981,Giustino_2010,Monserrat_2013,Han_2013,
Monserrat_Diamond_2014}
\begin{equation}\label{eq:expansion}
 E_{\text{g}}(\mathbf{q})= E_{\text{g}}(\mathbf{0})+
\sum_{n,\mathbf{k}}c_{n\mathbf{k}}^{(1)}q_{n\mathbf{k}}+
\sum_{\substack{n,\mathbf{k}\\n',\mathbf{k}'}}c_{n\mathbf{k}\,n'\mathbf{k}'}^{(2)}
q_{n\mathbf{k}}q_{n'\mathbf{k}'}\,, 
\end{equation}
where we have retained terms up to second order.
Within the harmonic approximation, the vibrational wave function is even and
the only non-zero terms in the expectation value of Eq.~(\ref{eq:band_gap})
using the expression given by Eq.~(\ref{eq:expansion}) are the quadratic
diagonal terms with coupling constants $c_{n\mathbf{k}\,n\mathbf{k}}^{(2)}$.
Within the BO approximation, the coupling constants are independent of
temperature and we therefore focus on the zero-point renormalization (ZPR) to
the band gap, which can be written as
\begin{equation}\label{eq:zpc}
E_{\text{ZPR}}=\sum_{n,\mathbf{k}}
\frac{c_{n\mathbf{k}\,n\mathbf{k}}^{(2)}}{2\omega_{n\mathbf{k}}}.
\end{equation}
This expression excludes terms with powers of $q_{n\mathbf{k}}$ higher than two
and any description of coupling between different vibrational modes, but it 
has been found to produce good agreement with experimental results for a range 
of materials~\cite{Antonius_2014,Giustino_2010,Han_2013}.

\subsection{Results}

We used the harmonic wave functions, obtained as described in 
Sec.~\ref{sec:latt_dyn}, to determine each $c_{n\mathbf{k}\,n\mathbf{k}}^{(2)}$ by 
performing frozen phonon calculations for each of the $\mathbf{k}$-points in 
the IBZ.
The frozen phonon calculations for the vibrational mode labelled 
($n,\mathbf{k}$) were performed using a vibrational amplitude of magnitude 
$\sqrt{\langle q_{n\mathbf{k}}^{2} \rangle}/2$, and we averaged over positive 
and negative displacements.

\begin{figure}
\centering
\includegraphics[width=\linewidth]{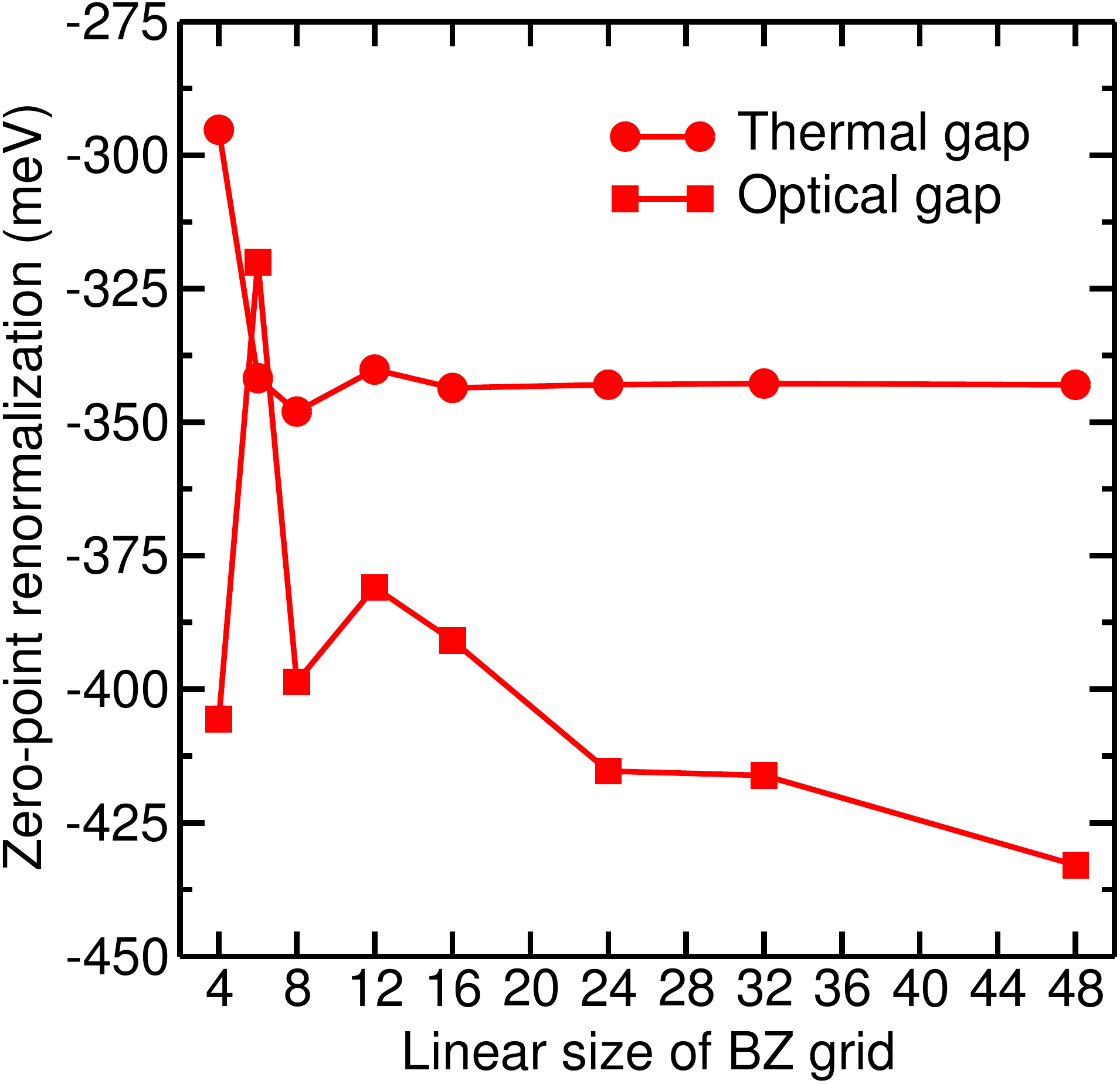}
\caption{(color online) ZPR to the thermal (red circles) and optical (red
squares) band gaps of diamond as a function of the linear size of the BZ grid. 
The solid lines are a guide to the eye.}
\label{fig:diamond_zpc}
\end{figure}

In Fig.~\ref{fig:diamond_zpc}, we show the ZPR to the thermal and optical band 
gaps of diamond as a function of the linear size of the BZ grid. 
It is well-known that these quantities converge slowly with respect to the 
number of points used to sample the vibrational BZ~\cite{Giustino_2010} and 
highly converged results have previously only been obtained using perturbative 
methods.
The largest grids explored so far using the direct method for diamond are of 
sizes $4 \times 4 \times 4$~\cite{Antonius_2014} and $6 \times 6 \times
6$~\cite{Monserrat_Diamond_2014}.
Here, we report results calculated using vibrational BZ grids of size up to $48
\times 48 \times 48$, vastly increasing the capabilities of the direct method 
for this type of calculation.
We find that the ZPR to the thermal gap converges within $1$~meV to a value of
$-343$~meV at a grid size of $24 \times 24 \times 24$.
The ZPR to the optical gap has a value of about $-430$~meV at a grid size of
$48 \times 48 \times 48$ and this value differs by $15$~meV from that 
calculated using a $32 \times 32 \times 32$ grid.

We have shown that we are able to use the direct method to calculate a value of 
the ZPR to the thermal band gap of diamond that is converged to better than 
$1$~meV with respect to the number of points used to sample the vibrational BZ. 
The ZPR to the optical gap converges more slowly, and the results from the 
largest grids we consider have an uncertainty about an order of magnitude
greater, of the order of $10$~meV. 
The choice of pseudopotential~\cite{Ponce_2014}, higher-order terms in 
Eq.~(\ref{eq:expansion})~\cite{Monserrat_Helium_2014}, and many-body 
effects~\cite{Antonius_2014} are known to change the values of the ZPR
by amounts greater than these levels of convergence.
The computational cost of investigating some of these effects may also
be greatly reduced by the use of non-diagonal supercells.

\section{Conclusions}\label{sec:conclusions}

We have described the use of non-diagonal supercells to study the response of 
periodic systems to perturbations characterized by a wave vector.
We have shown that, for a wave vector with reduced fractional coordinates 
$(m_1/n_1,m_2/n_2,m_3/n_3)$, there exists a commensurate supercell containing 
a number of primitive cells equal to the least common multiple of $n_1$, $n_2$, 
and $n_3$.
This compares favourably with the $n_1n_2n_3$ primitive cells required if 
only diagonal supercells are used.

We have compared the use of diagonal and non-diagonal supercells for
performing first principles lattice dynamics calculations using the direct
method.
We find over an order of magnitude reduction in the computational cost of
obtaining converged zero-point energies and phonon dispersions for diamond and 
graphite when using non-diagonal supercells.
We have also investigated the zero-point renormalization to the thermal and 
optical band gaps of diamond arising from electron-phonon coupling.
Utilizing non-diagonal supercells has allowed us to perform these calculations 
with Brillouin zone grids of sizes up to $48 \times 48 \times 48$. 
Our results show unprecedented levels of convergence for the values of the 
zero-point renormalization to the thermal and optical gaps calculated using the
direct method, of the orders of $1$~meV and $10$~meV, respectively.

The responses of condensed matter systems to perturbations characterized by a
wave vector are central in probing a wide range of physical properties, such as 
phonon dispersions~\cite{Giannozzi_1991}, electron-phonon 
coupling~\cite{Kong_2001}, spin fluctuations~\cite{Savrasov_1998}, nuclear 
magnetic resonance $J$-coupling~\cite{Joyce_2007}, and many-body dispersion 
effects~\cite{Ambrosetti_2014}.
Perturbative methods have provided a computationally efficient manner of 
determining these responses using first principles methods. 
The direct method has previously been considered computationally expensive due 
to the need to use simulation cells containing multiple primitive cells. 
However, it is more transparent, easier to implement in computer codes, and 
can be used in situations when it is necessary to go beyond the linear response
regime. 
The use of non-diagonal supercells described in this paper significantly 
reduces the computational cost of the direct method, and therefore expands its 
applicability to problems that were previously only tractable using 
perturbative methods. 

\begin{acknowledgments}
We thank Neil Drummond, Phil Hasnip, Miquel Monserrat, and Richard Needs for 
useful discussions, Tim Mueller for pointing us in the direction of 
Ref.~\onlinecite{Hart_2008}, and Michael Rutter for maintaining the 
{\sc check2xsf} program, which we used to construct the non-diagonal supercells.
J.\,H.\,L.-W.\ thanks the Engineering and Physical Sciences Research Council 
(UK) for a PhD studentship.
B.\,M.\ thanks Robinson College, Cambridge, and the Cambridge Philosophical 
Society for a Henslow Research Fellowship.
This work used the Cambridge High Performance Computing Service, for which
access was funded by the EPSRC [EP/J017639/1], and the ARCHER UK National
Supercomputing Service, for which access was obtained via the UKCP consortium
and funded by the EPSRC [EP/K013564/1].
\end{acknowledgments}

All relevant data present in this article can be accessed at:
\url{https://www.repository.cam.ac.uk/handle/1810/251429}.

\appendix*

\section{Complete and reduced residue systems}

Here we summarize some properties of complete and reduced residue systems.
Further details can be found in Ref.~\onlinecite{Number_Theory}.

If $a\,\text{mod}\,n=b\,\text{mod}\,n$, then $a$ is said to be 
\textit{congruent} to $b$ modulo $n$.

Numbers which are congruent modulo $n$ form an \textit{equivalence class} 
modulo $n$.

Any member of an equivalence class is said to be a \textit{residue} 
modulo $n$ with respect to all the members of the equivalence class.
Taking one residue from each equivalence class, we obtain a \textit{complete 
residue system} modulo $n$.

If the GCD of $a$ and $n$ is equal to unity, and $x$ runs over a complete
residue system modulo $n$, then $ax$ also runs over a complete residue system 
modulo $n$.
$a$ is therefore a generator for the additive group of integers modulo $n$.

The members of an equivalence class modulo $n$ all have the same GCD relative
to the modulus.
Taking one residue from each class for which the GCD relative to the modulus is 
equal to unity, we obtain a \textit{reduced residue system} modulo $n$.

If the GCD of $a$ and $n$ is equal to unity, and $x$ runs over a reduced
residue system modulo $n$, then $ax$ also runs over a reduced residue system 
modulo $n$.
$ax$ is therefore a generator for the additive group of integers modulo $n$.

\bibliography{supercell}

\end{document}